\documentclass[a4paper,usenatbib,useAMS]{mnras}
\usepackage{graphicx}	
\usepackage[flushleft]{threeparttable}
\usepackage{amsmath}	
\usepackage{amssymb}	
\usepackage{multicol}        
\usepackage{bm}		
\usepackage{pdflscape}	
\usepackage{color}
\usepackage{hyperref}
\usepackage{natbib}
\usepackage[T1]{fontenc}
\usepackage{ae,aecompl}
\usepackage{newtxtext,newtxmath}
\usepackage[utf8]{inputenc} 
 
\title[New lensed quasars in the Southern Sky]{Bright lenses are easy to find: \\ Spectroscopic confirmation of lensed quasars in the Southern Sky}

\author[C.~Spiniello et al.] {C.~Spiniello$^{1,2}$,  A.~Agnello$^{2,3}$, A.~V.~Sergeyev$^{4,5}$, T.~Anguita$^{6,7}$,  \'O. Rodr\'iguez$^{6,7}$, \and N.~R.~Napolitano$^{1,8}$,  C.~Tortora$^{9}$
\\
$^{1}$INAF - Osservatorio Astronomico di Capodimonte, Salita Moiariello, 16, I-80131 Napoli, Italy \\
$^{2}$European Southern Observatory, Karl-Schwarschild-Str. 2, 85748 Garching, Germany \\
$^{3}$Dark Cosmology Centre, Juliane Maries Vej 30, 2100 Copenhagen, Denmark\\
$^{4}$Astronomical Institute of Kharkov National University, 61022, 35 Sumskaya St, Kharkov, Ukraine \\
$^{5}$Institute of Radio Astronomy of the National Academy of Sciences of Ukraine \\
$^{6}$Departamento de Ciencias Fisicas, Universidad Andres Bello Fernandez Concha 700, Las Condes, Santiago, Chile \\
$^{7}$Millennium Institute of Astrophysics, Av. Vicuna Mackenna, 4860, Macul, Santiago, Chile \\
$^{8}$School of Physics and Astronomy,  Sun Yat-sen University Zhuhai Campus, 2 Daxue Road,  Tangjia,  Zhuhai,  Guangdong 519082,  P.R. China \\
$^{9}$INAF - Osservatorio Astrofisico di Arcetri, Largo Enrico Fermi 5, 50125, Firenze, Italy}

\date{Last updated 2018 December 06}

\pubyear{2018}
\begin{document}
\label{firstpage}
\pagerange{\pageref{firstpage}--\pageref{lastpage}}
\maketitle

\begin{abstract}
Gravitationally lensed quasars are valuable, but extremely rare, probes of observational cosmology and extragalactic astrophysics. Progress in these fields has been limited just by the paucity of systems with good ancillary data. 
Here we present a first spectroscopic confirmation of lenses discovered in the Southern Sky from the DES and KiDS-DR3 footprints. 
We have targeted 7 high-graded candidates, selected with new techniques,  with NTT-EFOSC2, and confirmed 5 of them. We provide source spectroscopic redshifts, image separations, $gri$ photometry and first lens model parameters. The success rate of $\sim 70\%$ confirms our forecasts, based on the comparison between the number of candidate doubles and quadruplets in our searches over a $\approx5000\text{ deg}^2$ footprint and theoretical predictions. 
\end{abstract}


\begingroup
\let\clearpage\relax
\endgroup
\newpage

\section{Introduction} 
\label{sec:intro}
When a quasar (QSO) is strongly lensed by a galaxy, it results in multiple images of the same source, possibly accompanied by arcs or rings that map the lensed host galaxy. The light-curves of different images are offset by a measurable time-delay that depends on cosmological distances to lens and source and the gravitational potential of the lens \citep{Refsdal64}, which in turn enables one-step measurements of the expansion history of the Universe and the dark matter halos of massive lens galaxies at up to z$\sim$0.5 \citep[e.g.][]{Suyu14}. 
The microlensing effect on the multiple QSO images, induced by stars in the deflector, provides a quantitative handle on the stellar content of the lens galaxies (e.g. \citealt{Schechter02, Bate11, Oguri14}) and, simultaneously, can constrain the inner structure of the source quasar, both accretion disk size and thermal profile (e.g. \citealt{Anguita08, Eigenbrod08, Motta12}) as well as the geometry of the broad line region (e.g. \citealt{Sluse11, Guerras13, Braibant14}). %
Finally, source reconstruction of the lensed QSO and its host give a direct view of QSO-host coevolution up to z$\sim2$ \citep[e.g.][]{pen06,din17}. 
However, lensed quasars are rare on the sky: typically $\sim0.1$ per $\text{ deg}^2$ at depth and resolution of present day surveys \citep[][ hereafter OM10]{Oguri10}, since they require a very close alignment of quasars with foreground massive galaxies, or galaxy clusters. 
The advent of various wide-field surveys, charting extended areas in both hemispheres, helps to overcome the intrinsic rarity of lensed quasars, provided that suitable techniques are devised in order to mine them in massive databases.

Within our team, we have developed multiple techniques of data-mining for lensed quasar searches 
and recently applied them with success to the Kilo Degree Survey and to the Dark Energy Survey (\citealt{Spiniello18}, and \citealt{Agnello18}, hereafter S18 and A18 respectively).
We combined different methods to pre-select QSOs
, tested their performance and complementarity, and published a list of high-grade candidates, to facilitate (possibly independent) spectroscopic follow-up campaigns. 

Here, we present results from the first of these follow-up campaigns, based on data acquired with EFOSC2 (the ESO Faint Object Spectrograph and Camera) mounted at the Nasmyth B focus of the 3.6m New Technology Telescope (NTT). 
We give details on the selected candidates in the next Section (Sec.~\ref{spec_obs}). We report on the spectroscopic run in Section~\ref{results} and provide photometry and lens model results in Section~\ref{models}. We finally conclude in Section~\ref{discuss}.\\
Whenever needed, we adopt a flat $\Lambda$CDM cosmology with $\Omega_{\Lambda}=0.7$ and $H_{0}=70$ km/s/Mpc. Magnitudes from DES and KiDS are in the AB system while the ones from WISE are in the Vega system.

\section{Candidate selection} 
\label{spec_obs}
We selected our candidates from S18, A18 and from a more recent (hitherto unpublished) list of high-grade candidates found in a similar way. 
We refer to the two mentioned papers for a detailed description of the morphology- and photometry-based methods that we developed to find lens candidates in wide-sky photometric multi-band Surveys. 
The new DES candidates were selected using astrometric offsets of WISE-preselected objects (with the same WISE cuts as in A18) between DES and Gaia J2000 coordinates, whereas in A18 we were considering offsets between 2MASS and Gaia coordinates. The reason behind this method can be explained by noting that if the deflector and quasar images contribute differently in different bands or in data with different image quality, this should result in centroid offsets of the same object among different surveys. 
We followed the same procedure as in A18 to detrend such astrometric offsets, and targets were required to have field-corrected offsets between $0.27\arcsec$ and $2.0\arcsec$ (following tests by S18 and A18).

Targets were then visually inspected via the NCSA-DESaccess cutout server\footnote{\texttt{https://des.ncsa.illinois.edu/easyweb}} to assemble a final list of candidates. Some high-grade candidates were lost at visual inspection, due to visualization issues from large queries in the NCSA web interface (we expected this since A18). As per STRIDES\footnote{STRIDES (\texttt{strides.astro.ucla.edu}) is a broad external collaboration of the DES.} internal agreements, the full list of lens candidates from DES-vs-Gaia offsets will not be disclosed yet. 

Some of our candidates have also been found by another group at the same time (Lemon et al. in prep.). Following publication agreements within STRIDES, we give here partially blinded coordinates for two of them.  
Table~1 reports coordinates and infrared magnitudes (used in the stage of pre-selection) of the systems. 

\begin{figure}
\begin{center}
\includegraphics[scale=0.3,angle=0]{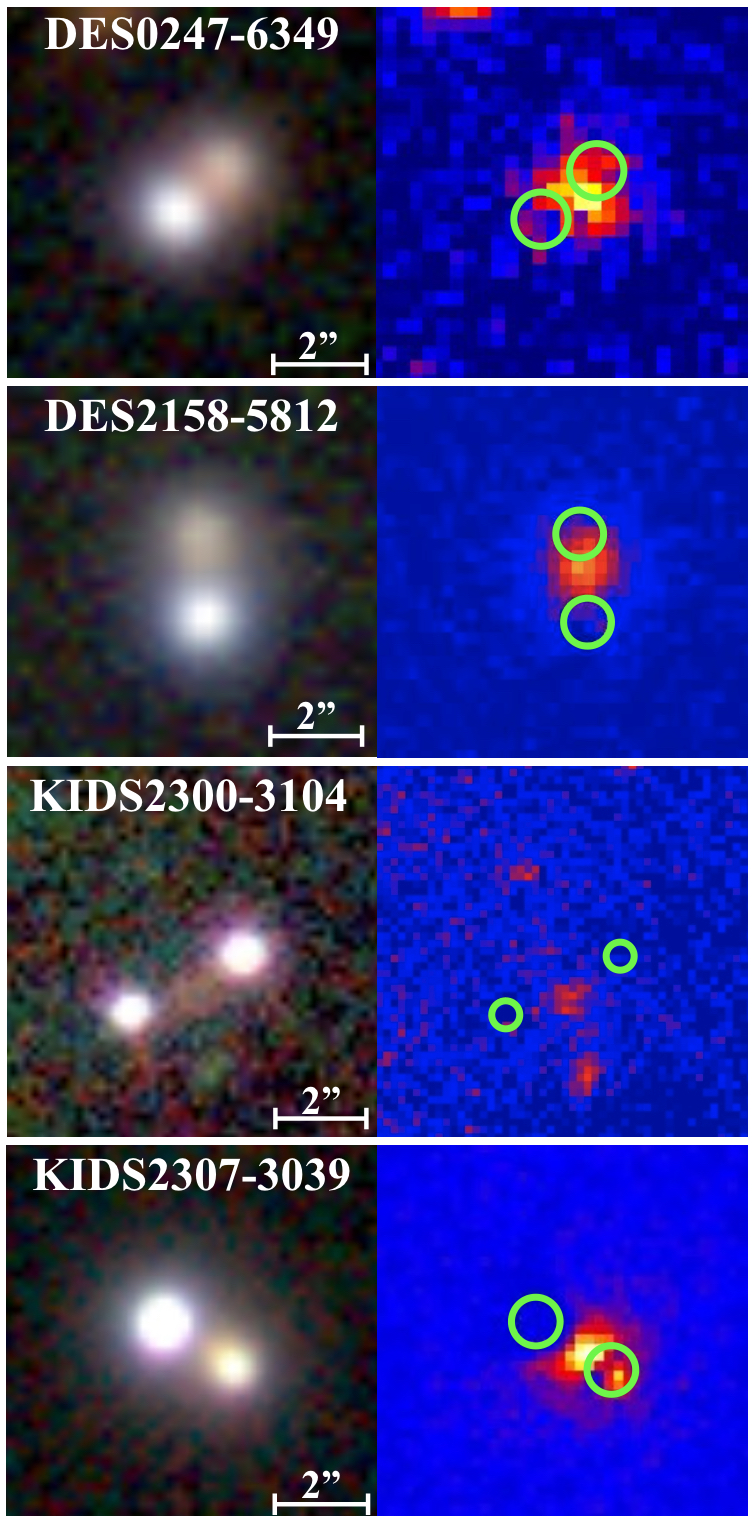}
\caption{Cutouts of $7\arcsec \times 7\arcsec$ in size generated combining $g$-, $r$- and $i$-band images (left panels) and DIA residuals (right panels) for the four candidates from A18 and S18. The cutouts are modeled as two-point sources plus a galaxy, and the two best-fit point sources are then subtracted.}
\label{fig:dia1}
\end{center}
\end{figure}

\begin{figure}
\begin{center}
\includegraphics[scale=0.27,angle=0]{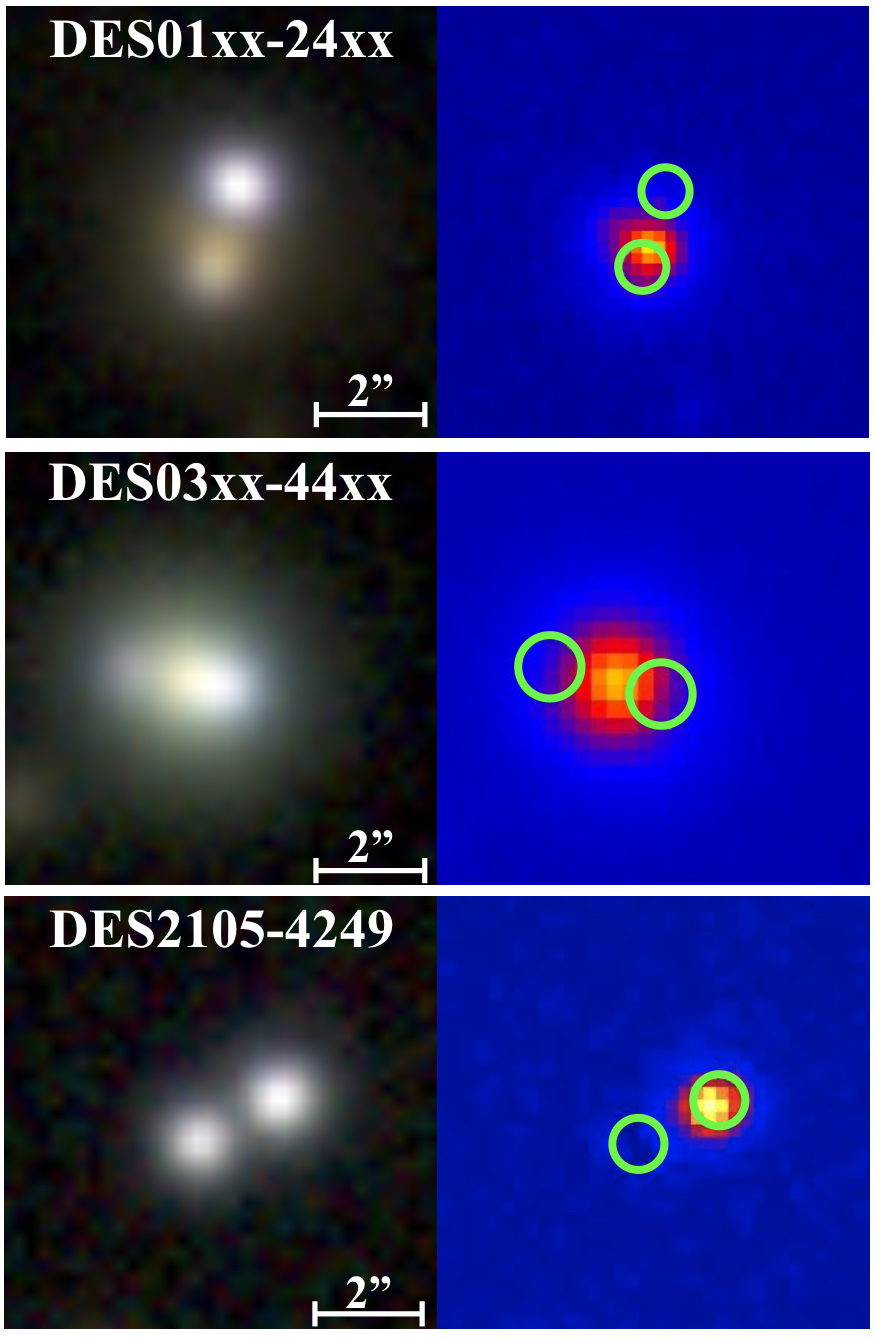}
\caption{$gri$-cutouts and DIA residuals (same as in Fig.~\ref{fig:dia1}) for the new candidates presented in this paper.}
\label{fig:dia2}
\end{center}
\end{figure}

\begin{table*}
\caption{List of observed candidates with the infrared magnitudes from the WISE catalog \citep{Wright10}, that have been used at the pre-selection stage (for some systems, errors of magnitudes are missing from the WISE catalog). Lenses with partly blinded coordinates have been found independently also by Lemon et al (in prep.), here indicated as L18. }
\label{tab:tab1}
\begin{center}
\begin{tabular}{|c|c|c|c|c|c|c|c|}
\hline 
ID & RA (J2000) & Dec (J2000) & W1 (mag) & W2 (mag)  & W3 (mag)  & W4 (mag) & Reference \\ 
\hline 
DES~0247-6349 & 02:47:54.77 & -63:49:23.20 & $15.10\pm0.03$ & $14.4\pm0.04$ & $11.4\pm0.1$ & $8.6\pm0.2$ &  A18 \\  
DES~2158-5812 & 21:58:37.30 & -58:12:03.90 & $14.70\pm0.03$ & $13.9\pm0.03$ & $11.5\pm0.2$ & $8.5\pm0.3$ &  A18 \\ 
KIDS~2300-3104 & 23:00:11.82 & -31:04:07.07 & $16.48\pm0.09$ & $15.7\pm0.15$ & 11.8 & $9.1\pm0.5$ &  S18 \\ 
KIDS~2307-3039 & 23:07:18.87 & -30:39:15.96 & $14.88\pm0.03$ & $14.0\pm0.04$ & $10.8\pm0.1$ & 8.2 &  S18 \\ 
DES~2105-4249 & 21:05:04.20 & -42:49:36.60 & $15.59\pm0.04$ & $14.3\pm0.05$ & $11.6\pm0.2$ & 8.4 &  this paper\\
DES~01xx-24xx & 01:XX:XX.X & -24:XX:XX.X & $14.15\pm0.03$ & $13.17\pm0.03$ & $10.38\pm0.08$ & $7.5\pm0.1$ & this paper, and L18 \\ 
DES~03xx-44xx & 03:XX:XX.X & -44:XX:XX.X & $14.56\pm0.03$ & $14.2\pm0.03$ & $12.1\pm0.2$ & 8.9 & this paper, and L18 \\ 
\hline 
\end{tabular} 
\end{center}
\end{table*}

For all the candidates, we perform Direct Image Analysis (DIA), as detailed in S18. Briefly, a PSF model is fit and subsequently subtracted from the survey image to check for the presence of signal from the deflector. In particular, using the \texttt{PyRAF} package\footnote{PyRAF is a product of the Space Telescope Science Institute, which is operated by AURA for NASA.} on the $r-$band images, we simultaneously fit a point-PSF model to the QSO multiple images.  We then generated subtracted images which we visually inspect to identify the  position of the deflector. 
For the DES systems, DIA was performed after the spectroscopic run, since we did not use it in A18.

We show the $gri$-cutouts (left panel) and the DIA residuals (right panels) for the candidates reported in A18 and S18 in Figure~\ref{fig:dia1}, and for the new candidates presented in this paper in Figure~\ref{fig:dia2}.

\section{Spectroscopic follow-up} 
\label{results}
Three nights on NTT-EFOSC2 (P0101.A-0298, PI: Anguita) were allocated for follow-up in July 2018. However, the run was mostly weathered out and we could observe only 7 candidates, each with 20 minutes of integration time. The seeing was stable at $1.0\arcsec\pm0.2\arcsec,$ spectra were reduced using standard (ESO-released) pipelines, and the quasar traces were optimally de-blended using a superpostion of Gaussian profiles in the spatial direction \citep[see][]{mor04, agn15b}.

We have confirmed 5 out of 7 lenses, selected from wide-sky photometric Surveys, namely DES and KiDS, in the Southern Sky. Concerning the two remaining candidates, KIDS2300-3104 resulted to be a couple of stars, while for bf DES2105-4249 only one trace showed emission lines in its spectrum. This system is therefore not a lensed quasar where the two point-like sources are multiple images of the same object. It is most likely a projection effect of a galaxy+QSOs.  

This success rate of $\sim70\%$ is slightly higher but still consistent with the estimates from A18, considering the number of high-graded candidates versus the number of expected lenses over the DES footprint (adopting the lensing rates by OM10). 

\begin{figure*}
\begin{center}
\includegraphics[scale=0.22,angle=0]{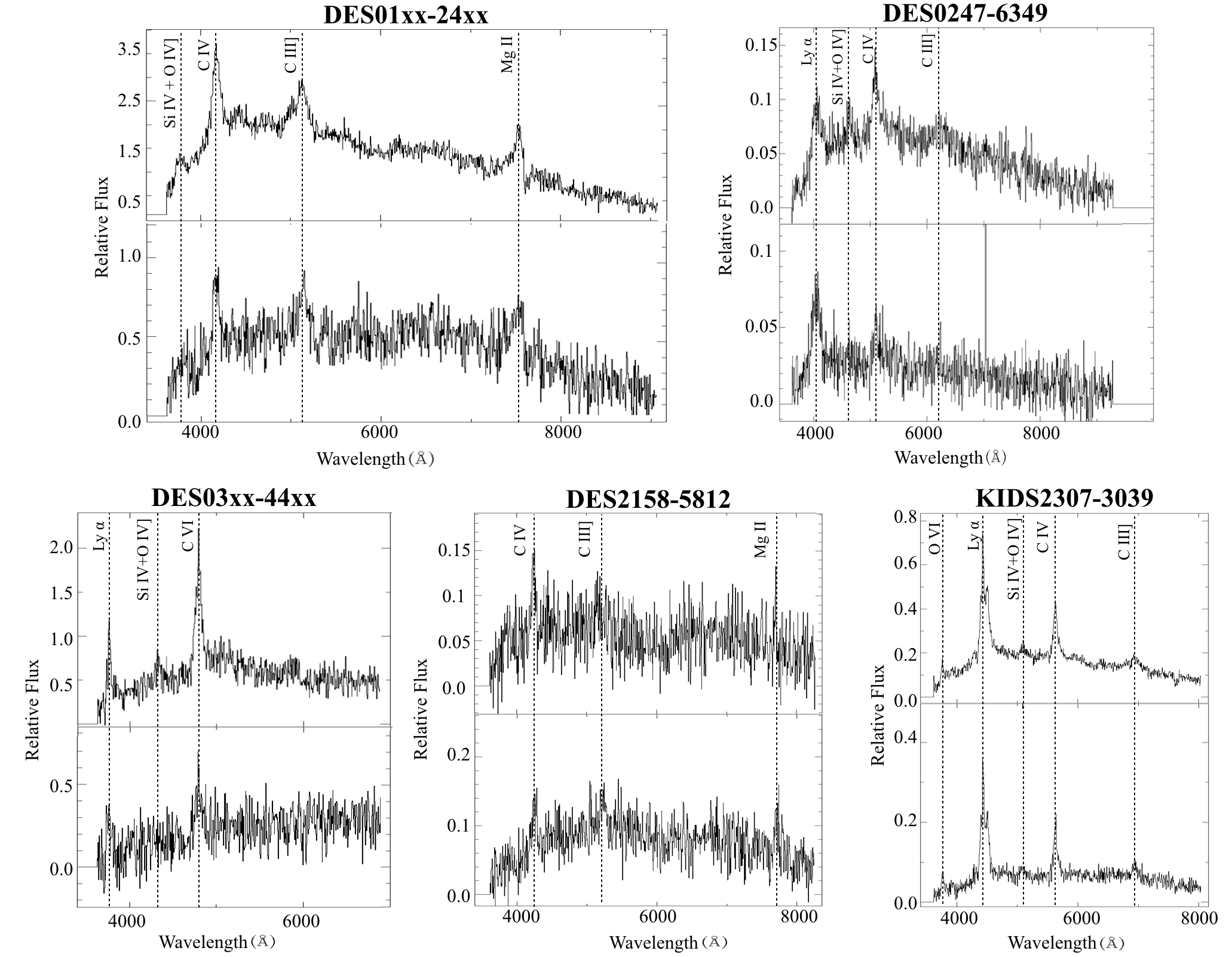}
\caption{1D extracted spectra of the two quasar images for the five confirmed new lenses. The most prominent emission lines which allowed us to infer the redshift of the sources are highlighted in the plots. Component A (brighter) is always plotted in the upper panels, while the component B (fainter) in the lower ones. The wavelength is always plotted in the observed frame. }
\label{fig:spectra}
\end{center}
\end{figure*}

The NTT spectra of the five confirmed lensed QSOs are shown in Figure~\ref{fig:spectra}, where we also highlight some of the stronger emission lines that enabled the determination of source redshifts, such as CIV~$\lambda1549$, L$\alpha$ or MgII~$\lambda2799$.  
The redshifts were determined with a simultaneous Gaussian fitting to the main emission lines on the spectra of the single components of the quasars, after subtracting the continuum with a 4th order polynomial function.  
We note that, although some features deviate from a Gaussian profile, this assumption is sufficient to correctly fit the peak of the lines and measure the redshift of their central wavelength.  Once identified the $i-$th emission lines and their rest frame wavelength ($\lambda_{0, \rm i}$), we defined the peaks of the Gaussians  as $\lambda_{0, \rm i}(1+z)$, hence $z$ is a common free parameter for the fitting function, defined as:  
\begin{equation}
\Sigma_{i} A_i e^\frac{[(\lambda-\lambda_{0, \rm i})(1+z)]^2}{2\sigma_i^2}
\end{equation}
where $A_i$ are normalisation factors which account for the peak high, and $\sigma_i$ measure the width of the line.  
In order to assess the statistical error on the redshift estimates, we have re-sampled 100 times the spectrum of every source in the fitted range including the three brightest emission line features. We have randomly re-extracted each pixel value around the current one assuming Gaussian noise given by the rms of the continuum in the regions close to the fitting window. We have then obtained the mean and standard deviation of the best-fit redshift obtained using a Levenberg-Marquardt method for $\chi^2$ minimization over the 100 re-sampled spectra. These are reported in Table~\ref{tab:table_res} for all confirmed quasars as final redshift estimates with their statistical uncertainty. 


Unfortunately, the faint traces of the lensing galaxies in the spectra did not allow us to securely identify deflector redshifts using the same method. We then estimate photometric redshift distributions using deflector magnitudes from the DES and KIDS multi-band images, as explained below.

Using \texttt{galfit} \citep{Peng02} we performed PSF photometry on all the components (A = brightest QSO image, B = fainter QSO image, G = lens galaxy) for $g$-, $r$-, and $i$-bands. We use a Se\'rsic profile with n=4 (i.e. we assume a de Vaucouleurs profile) to fit the light distribution of the deflector and point-PSF to fit that of each QSO image. We fit all the components simultaneously.
We include the results in Table~2, together with the spectroscopic redshift values for the sources and the photometric redshift values for the deflectors. 

\begin{figure}
\begin{center}
\includegraphics[scale=0.12,angle=0]{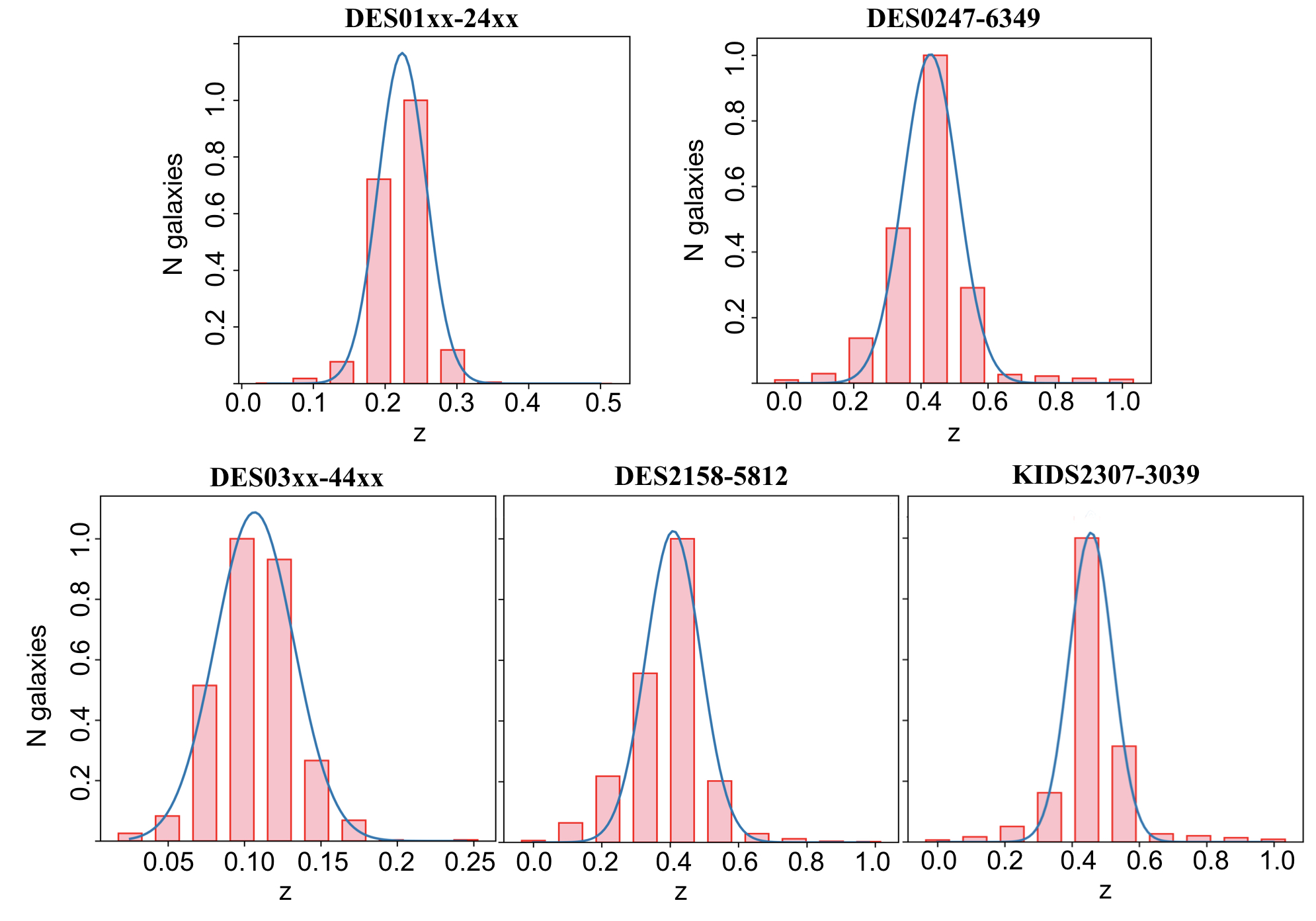}
\caption{Inference on the photometric redshifts of the deflectors in this sample, using spectroscopic redshifts from SDSS-DR14 \texttt{SpecPhotoAll} galaxies with similar PSF colours (within 3$\sigma$ uncertainties). The number of galaxies plotted on the y-axis for each system has been normalized to the histogram maximum value.}
\label{fig:histog}
\end{center}
\end{figure}

To estimate the photometric redshifts of the deflectors we compared their PSF colors as derived above with galaxy spectra from the Sloan Digital Sky Survey (SDSS, DR14, \citealt{Abolfathi18}). In particular, for each deflector, we select all galaxies in \texttt{SpecPhotoAll} with colours within 3$\sigma$ from the ones obtained from PSF photometry (we use the \texttt{psfMag} and its \texttt{psfMagErr} for each band) on the lens and retain their spectroscopic redshifts. Then, we fit a normal distribution to the resulting histograms (plotted in Fig.~\ref{fig:histog}), obtaining the mean and standard deviation of the redshift, that we adopted as the fiduciary photo-$z$ and its uncertainty. 

\begin{table*}
\caption{Astrometric properties of all the components of the five confirmed lenses. Relative positions are given in arcsec and always using as reference the brightest QSOs image (A). Magnitudes in the $gri$ bands have been computed performing PSF photometry, using \texttt{galfit} (see Sec.~\ref{results}).}
\label{tab:table_res}
\begin{center}
\begin{tabular}{|c|c|c|c|c|c|c|c|}
\hline 
ID &  comp. & $\delta$x & $\delta$y & $z$ & $g$ & $r$ & $i$  \\ 
   &            & (arcsec)  &  (arcsec) & (redshift) & (mag) & (mag) & (mag) \\
\hline 
DES01xx-24xx & A & 0.000$\pm$0.003 & 0.000$\pm$0.003  & 1.692$\pm$0.001  & 18.83$\pm$0.01  & 18.87$\pm$0.01 & 18.62$\pm$0.01 \\ 
DES01xx-24xx & B & -0.523$\pm$0.013 & -1.770$\pm$0.013 & 1.690$\pm$0.002  & 20.49$\pm$0.02 & 20.25$\pm$0.05 & 19.91$\pm$0.07 \\ 
DES01xx-24xx & G & -0.434$\pm$0.024 & -1.283$\pm$0.037  & 0.22$\pm$0.03 & 20.30$\pm$0.04  & 18.87$\pm$0.03  & 18.31$\pm$0.02  \\ 
\hline 
DES0247-6349 & A & 0.0$\pm$0.003  & 0.0$\pm$0.005  & 2.303$\pm$0.009  & 19.95$\pm$0.01  & 19.80$\pm$0.02  & 19.81$\pm$0.03 \\ 
DES0247-6349 & B & +1.055$\pm$0.013  & 0.984$\pm$0.013  & 2.31$\pm$0.02 &  21.13$\pm$0.02  & 20.87$\pm$0.06  & 20.45$\pm$0.09 \\ 
DES0247-6349 & G & +0.76$\pm$0.19  & 0.60$\pm$0.16  & 0.43$\pm$0.08 & 22.21$\pm$0.23  & 21.11$\pm$0.17  & 19.87$\pm$0.13 \\ 
\hline 
DES03xx-44xx & A & 0.000$\pm$0.008  & 0.000$\pm$0.011  & 2.094$\pm$0.001  & 19.91$\pm$0.02  & 19.88$\pm$0.03  & 19.56$\pm$0.02 \\ 
DES03xx-44xx & B & -1.89$\pm$0.03  & 0.44$\pm$0.03  & 2.093$\pm$0.003  & 21.06$\pm$0.02  & 21.05$\pm$0.05  & 20.64$\pm$0.03  \\ 
DES03xx-44xx & G & -0.742$\pm$0.011  & 0.179$\pm$0.005  & 0.11$\pm$0.03  & 19.08$\pm$0.01  & 18.11$\pm$0.01  & 17.70$\pm$0.01 \\ 
\hline 
DES2158-5812 & A & 0.000$\pm$0.005  & 0.000$\pm$0.005  & 1.756$\pm$0.004  & 20.00$\pm$0.01  & 19.92$\pm$0.01  & 19.71$\pm$0.01 \\ 
DES2158-5812 & B & -0.124$\pm$0.016  & 1.888$\pm$0.018  & 1.74$\pm$0.02  & 21.71$\pm$0.01  & 21.26$\pm$0.10  & 21.09$\pm$0.18  \\ 
DES2158-5812 & G & -0.03$\pm$0.04  & 1.29$\pm$0.07  & 0.41$\pm$0.08  & 21.65$\pm$0.14  & 20.53$\pm$0.05  & 19.64$\pm$0.08 \\ 
\hline 
KIDS2307-3039 & A & 0.000$\pm$0.008  & 0.000$\pm$0.008  & 2.640$\pm$0.005  & 18.64$\pm$0.02  & 18.49$\pm$0.02  & 18.57$\pm$0.04 \\ 
KIDS2307-3039 & B & -2.046$\pm$0.013  & -1.291$\pm$0.016  & 2.641$\pm$0.003  & 20.01$\pm$0.04  & 19.78$\pm$0.04  & 19.18$\pm$0.07 \\ 
KIDS2307-3039 & G & -1.56$\pm$0.13  & -0.90$\pm$0.13  & 0.46$\pm$0.06  & 21.6$\pm$0.3  & 20.97$\pm$0.11  & 19.85$\pm$0.07 \\ 
\hline 
\end{tabular} 
\end{center}
\end{table*}

\section{Lens Models}
\label{models}
We provide here lens model parameters for the confirmed systems. We adopt a Singular Isothermal Sphere with external shear (SIS+XS), described by an Einstein radius $R_{Ein}$ and a shear amplitude $\gamma$ and orientation. The models are fit to the relative astrometry of quasar images and deflectors from above. The errors are propagated from the shear direction and are proportional to$1\pm \gamma$. The resulting Einstein radii are approximately half the image separations (within $\approx10\%$), but the SIS+XS model allows us to account for degeneracies between monopole and quadrupole contributions to the mass. 
The resulting best-fit parameters are given in Table~\ref{tab:table_model}, where we also give the predicted magnitude differences and time-delays. The main contribution on the time-delay uncertainties comes from the photometric redshift estimation. 
The model magnitude differences are in perfect agreement with the measured magnitude difference in $i$-band reported in Table~2 for DES01xx-24xx, DES0247-6349 and KIDS2307-3039. For the remaining two systems, in one case the difference is underestimated by the model (DES03xx-44xx) and in the other case (DES2158-5812) it is overestimated by it. However, we note that multiple factors might affect the actual flux ratios of lensed quasars: the combination of time-delays and source-variability, microlensing, substructures, and differential dust-reddening in the lens plane. Given that generally the B images show the most chromatic variation (redder than the brightest images), extinction probably plays a major role. Only deeper spectra and  deeper photometry would enable quantitative studies of the objects (e.g. \citealt{Yonehara98,Agnello17}). 

Finally, we calculated the total projected mass within the Einstein radius, using the classical lens equation formula (see e.g. \citealt{Schneider92}): 
\begin{equation}
M_{Ein} = R_{Ein}^{2}\dfrac{c^2D_{l}D_{s}}{4GD_{ls}}
\end{equation}
where $D_{l}$ is the distance of the lens from the observer, $D_{s}$ is the distance of the source from the observer and $D_{ls}$ is the relative distance between the lens and the source. We include the masses in Table~\ref{tab:table_model}

\begin{table}
\caption{Best-fit SIS+XS model parameters of the confirmed lenses.}
\label{tab:table_model}
\begin{tabular}{|c|c|c|c|c|c|}
\hline
ID & $R_{Ein}$ & $M_{Ein} $ & $\gamma$ & $\mu^a$ & $\Delta$t$^b$\\
& $(")$ & $(10^{11}M_{\odot})$ &  & (mag) & (day) \\
\hline
DES 01xx-24xx & 0.97 &  1.04 & 0.06 & -1.30 & $25\pm 4$ \\
DES 0247-6349 & 0.72 & 1.05 & 0.04 & -0.73 & $23\pm 6$ \\
DES 03xx-44xx & 0.96 & 5.12 & 0.01 & -0.61 & $-6\pm 2$ \\
DES 2158-5812 & 0.80 & 1.32 & 0.16 & -2.35 & $43 \pm 13$ \\
KIDS 2307-3039 & 1.19 & 2.96 & 0.04 & -0.60 & $99 \pm 17$ \\
\hline
\end{tabular}
\begin{tablenotes}
\item {{\sl a}:  model magnitude difference between A and B images  } 
\item {{\sl b}: the errors are propagated from the shear direction uncertainties and are proportional to $1\pm\gamma$ } 
\end{tablenotes}
\end{table}

%

\section{Discussion and future prospects} 
\label{discuss}
We have presented a sample of previously unknown lenses, found in the Southern Hemisphere through new techniques relying on astrometric offsets, and spectroscopically confirmed.
The success rate of 5/7 meets the expectations from A18 and S18. 
The range of source-redshifts and Einstein radii is consistent with OM10 predictions, further suggesting that these searches are complete within the photometric completeness of Gaia.

\begin{figure}
\begin{center}
\includegraphics[scale=0.11,angle=0]{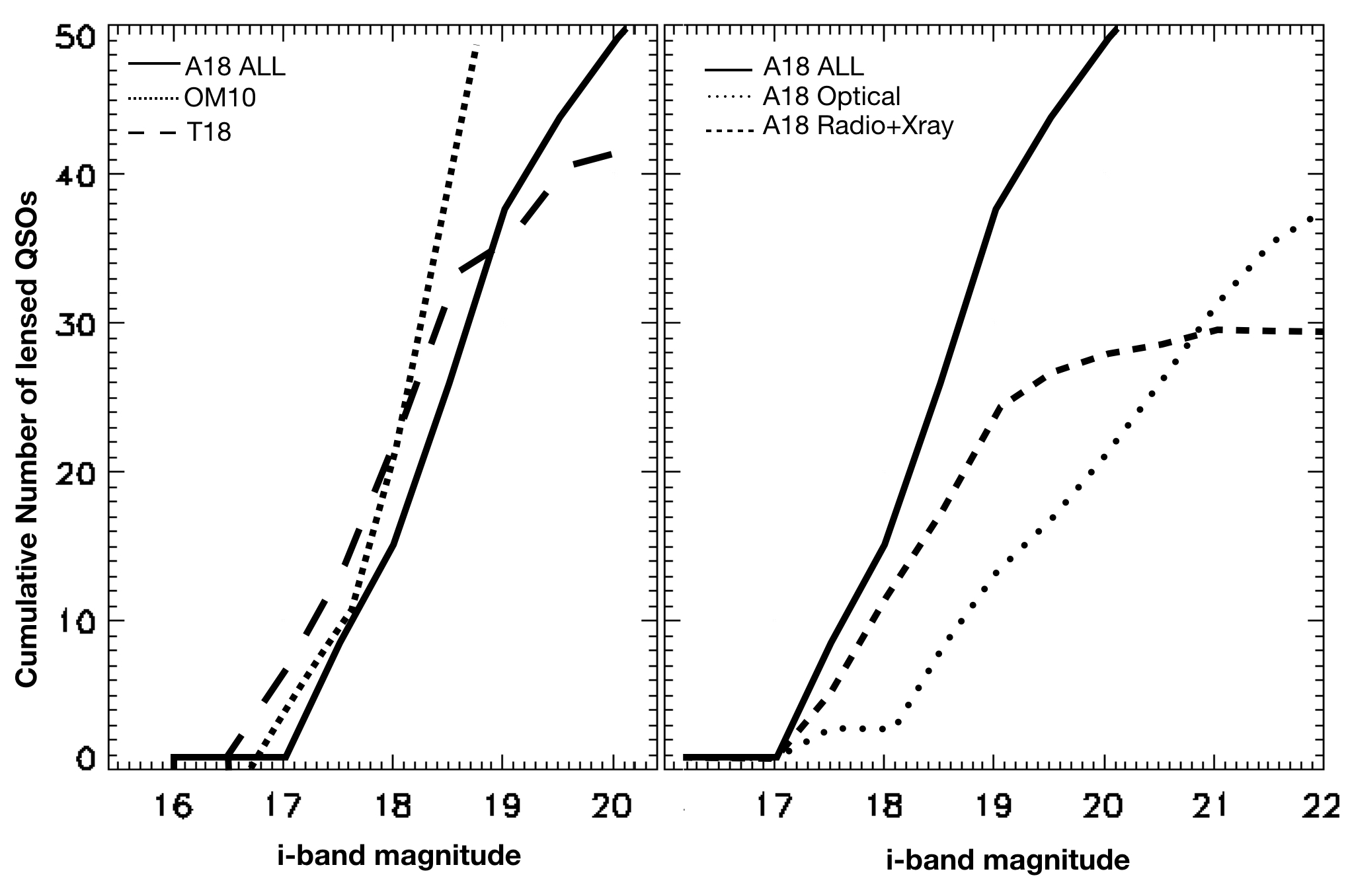}
\caption{{\sl Left panel:} Cumulative number of targets in bins of $i$-band magnitudes from A18 (solid lines) and scaled for the 70\% successful rate obtained here. The comparison with the distribution of targets in DES (dashed line) calculated by the STRIDES Collaboration (\citealt{treu18}) and with predictions from OM10 (dotted line) shows a very good agreement. {\sl Right panel:} Same as in the left panels, but the different lines show candidates selected in A18 from optical bands versus candidates selected from other wavelengths. Selecting QSOs from radio or X-Ray allows us to find fainter lenses.}
\label{fig:cumulative}
\end{center}
\end{figure}

Compared to the depth of DES and KiDS, these objects are still at the bright end.  This fact has been noted for various independent searches \citep[see][for forecasts on the DES]{treu18}: a large fraction of lenses with faint quasar sources is hitherto undiscovered, and different strategies of target and candidate selection are needed\footnote{See e.g. A18 for the discussion of radio and X-ray searches for lensed quasars over the DES footprint.}. This is shown in Figure~\ref{fig:cumulative}, where the cumulative distributions of predicted, candidate and confirmed lenses are shown. 
On the left panel, we show the distribution of targets in DES (dashed line) calculated by the STRIDES Collaboration (\citealt{treu18}) and the predictions from OM10 (dotted line). We compare these distributions with the ones of the A18 candidates, rescaled by $0.7$ to account for the success rate of this campaign, finding a very good agreement. 
Moreover, we note that samples relying on optical selection through Gaia, such as the one presented in \citet{treu18},  saturate between $i\approx$18.5 and $i\approx$20.0 whereas samples including radio and X-ray preselection (such as A18) extend to fainter magnitudes and follow the OM10 predictions more closely. The last point appears even more clearly in the right panel of Figure~\ref{fig:cumulative} where we split the A18 candidates according to the preselection method. Below $i\approx$20.5 the number of lens candidates found through radio and X-rays is larger than the number of systems selected in the optical. 

The brightness of most confirmed lenses also means that spectroscopic follow-up, to measure the kinematics of the deflectors, is affordable with current-generation facilities.
Through monitoring with nightly-cadence, an accuracy of $5\%$ should be attainable on the time-delay ($\Delta t$) between quasar lightcurves in systems with $\Delta t\gtrsim25$~days (Courbin, private comm.). 
Moreover, if the deflector's velocity dispersion $\sigma$ is measured, the combination $c^{3}\Delta t/\sigma^{2}\propto D_{l}$ of time-delay and kinematics would make these systems standard rulers at $z=0.2-0.65,$ to within $\approx20\%$ uncertainties \citep{Jee15,Shajib18}. 
For lenses with lower deflector-redshift, the time-delay distance $D_{\Delta t}=D_{l}D_{s}/((1+z_{l})D_{ls})$ is proportional to $D_l$ with a factor that is only weakly dependent on cosmological parameters ($\Omega_{\Lambda},$ $\Omega_{m}$). Securing spectroscopic redshifts of the deflectors and measuring time-delays would then yield two angular-diameter distances per lens.

\section*{Acknowledgment} 
The authors wish to thank the anonymous referee for interesting and constructive comments which helped in improving the quality and clarity of the paper. 

CS and NRN have received funding from the European Union's Horizon 2020 research and innovation programme under the Marie Sklodowska-Curie actions grant agreement No 664931 and No 721463 to the SUNDIAL ITN network, respectively. 
TA and OR acknowledge support by the Ministry for the Economy, Development and Tourism's Programa Inicativa Cientifica Milenio through grant IC120009, awarded to The Millennium Institute of Astrophysics (MAS).
CT acknowledges funding from the INAF PRIN-SKA 2017 program No 1.05.01.88.04.
%







\bsp	
\label{lastpage}
\end{document}